\documentclass[amsmath,11pt]{article}
\usepackage{amsfonts,amsmath,amssymb,graphics,epsfig}
\usepackage{enumerate}\usepackage{theorem}

\begin{document}

\title{\bf Raychaudhuri's equation and aspects of relativistic charged collapse}

\author{Alexandros P. Kouretsis and Christos G. Tsagas\\ {\small Section of Astrophysics, Astronomy and Mechanics, Department of Physics}\\ {\small Aristotle University of Thessaloniki, Thessaloniki 54124, Greece}}

\date{\empty}

\maketitle

\begin{abstract}
We use the Raychaudhuri equation to probe certain aspects related to the gravitational collapse of a charged medium. The aim is to identify the stresses the Maxwell field exerts on the fluid and discuss their potential implications. Particular attention is given to those stresses that resist contraction. After looking at the general case, we consider the two opposite limits of poor and high electrical conductivity. In the former there are electric fields but no currents, while in the latter the situation is reversed. When the conductivity is low, we find that the main agents acting against the collapse are the Coulomb forces triggered by the presence of an excess charge. At the ideal Magnetohydrodynamic (MHD) limit, on the other hand, the strongest resistance seems to come from the tension of the magnetic forcelines. In either case, we discuss whether and how the aforementioned resisting stresses may halt the contraction and provide a set of conditions making this likely to happen.
\end{abstract}

\section{Introduction}\label{sI}
Gravitational collapse has been one of the main topics of research interest in relativistic astrophysics for many decades. The implications of nonzero charges, together with those of electromagnetism, for the ultimate fate of a self-gravitating massive body, also have a long research history. Over the last fifty years, or so, several authors have looked into this question from a variety of perspectives and by means of different techniques. Some of the available work addresses static problems, while other authors have considered dynamical situations. For a representative list the reader is referred to~\cite{B}-\cite{LZ}, with more references given in~\cite{PHDMcLS}. A number of these studies have raised the possibility that a contracting charged medium may `bounce back' instead of `falling' into a singularity. The first signs were probably those seen in Melvin's `magnetic universe' and in the studies of collapsing charged dust by Novikov~\cite{M1}-\cite{N}. With very few exceptions, all of the aforementioned work uses metric-based techniques.

An alternative approach to gravitational collapse is to employ covariant methods and thus involve the Raychaudhuri equation~\cite{R2}. The latter determines the average contraction/expansion of a self-gravitating medium, by monitoring the cross-sectional area of the particle worldlines. Raychaudhuri's formula is the key equation of gravitational collapse and has played a central role in the development of the singularity theorems of the late 1960s and early 1970s (for recent reviews and extensive discussion the reader is referred to~\cite{E}-\cite{KS}). A major difference in the nature of the Raychaudhuri equation, when applied to a charged environment, is that the worldlines of the matter are no longer geodesics. This happens every time non-gravitational forces are included into the system and the particles' motion is not dictated by gravity alone. In our case, the fluid kinematics are determined by the combined action of the Einstein and the Maxwell fields. As a result, new stresses are added to Raychaudhuri's formula. These are primarily due to the Lorentz force and depend on the electric properties of the charged medium. On whether, in particular, the conductivity is low or high. In the first case, there are no electric currents despite the presence of a finite electric field. At the ideal Magnetohydrodynamic (MHD) limit, on the other hand, the situation is reversed.

For our purposes the stresses of interest are those that tend to resist contraction. When dealing with poorly conductive media, we find that the main resistance comes from the familiar Coulomb forces    triggered by the presence of a net charge. The overall effect, which is partly Newtonian and partly relativistic, depends on the specific charge of the matter. When the latter exceeds a particular threshold, the Coulomb stresses become strong enough to outbalance the gravitational pull of the matter fields and bring the collapse to a halt (see \S~\ref{ssPCD},~\ref{ssCRvsGA}). Analogous dynamical results, both qualitatively and quantitatively, have also been obtained through semi-covariant and metric-based methods~\cite{R1,KB}. When the conductivity is high the electric fields vanish and the currents keep the magnetic forcelines frozen-in with the fluid. Then, it appears that the main resistance to the gravitational pull of the matter fields comes from the magnetic tension. The latter reflects the elasticity of the field lines and their tendency to remain straight. This tendency is manifested through a variety of tension stresses that try to restore equilibrium every time the magnetic forcelines are distorted.

General Relativity describes the macroscopic behaviour of the gravitational field. In the geometrical framework of the theory, gravity is represented by the curvature of the spacetime, while the interplay between matter and geometry is monitored by Einstein's equations. However, for energy sources of vector nature, like the electromagnetic field for example, an additional set of equations is of almost equal importance. These are the Ricci identities. Our analysis shows that, within framework of General Relativity, one of the agents causing magnetic-line deformations is the curvature of the space. The mathematics of the interaction are monitored by means of the Ricci identities~\cite{T1}. As the collapse proceeds and the departure from the Euclidean flatness increases, the aforementioned deformations grow as well. As a result, the restoring tension stresses also get stronger. This opens the theoretical possibility that, when the curvature distortion crosses a certain threshold, the resulting tension stress may become strong enough to stop the contraction and, perhaps, turn it into expansion~\cite{T1}. Just like a spring bursts forward after been compressed back to its limit. We revisit this scenario in~\S~\ref{ssHCD},~\ref{ssMTEs} and show that some degree of anisotropy would probably be necessary, if the magnetic tension is to bring the collapse to a halt. The required amount of anisotropy is found to depend on the $B$-field's relative strength. We attempt to quantify these statements by providing the necessary conditions for the tension stresses to counterbalance the gravitational pull of the matter in Raychaudhuri's equation.

\section{The 1+3 covariant approach}\label{s1+3CA}
The covariant approach to general relativity utilizes an 1+3 `threading' of the spacetime, as opposed to the 3+1 `slicing' of the ADM decomposition (see~\cite{TCM} for a recent review). This is achieved by introducing the timelike 4-velocity $u^a={\rm d}x^a/{\rm d}\tau$, normalised so that $u_au^a=-1$. The $u_a$-field defines the worldlines of the fundamental observers (with $\tau$ being the associated proper time), while the tensor $h_{ab}=g_{ab}+u_au_b$ (with $g_{ab}$ representing the spacetime metric) projects orthogonal to $u_a$ and into the observers' 3-dimensional, instantaneous rest-space. On using $u_a$ and $h_{ab}$, every variable, operator and equation spits relative to the $u_a$-field into their irreducible timelike and spacelike components.

\subsection{The gravitational field}\label{ssGF}
Within the geometrical interpretation of general relativity, the gravitational field is described by means of the Riemann curvature tensor. The latter splits into a local (Ricci) and a long-range (Weyl) part according to
\begin{equation}
R_{abcd}= C_{abcd}+ {1\over2}\left(g_{ac}R_{bd}+g_{bd}R_{ac}-g_{bc}R_{ad}-g_{ad}R_{bc}\right)- {1\over6}R\left(g_{ac}g_{bd}-g_{ad}g_{bc}\right)\,,  \label{Riemann}
\end{equation}
where $C_{abcd}$ is the (conformal curvature) Weyl tensor and $R_{ab}$ is the Ricci tensor (with $R=R^a{}_a$ being the associated Ricci scalar). The Riemann tensor also obeys the symmetries $R_{abcd}=R_{[ab][cd]}=R_{cdab}$.

The Ricci field is determined by the local matter distribution and the interplay between geometry and matter is governed by the Einstein field equations. In the absence of a cosmological constant the latter read
\begin{equation}
R_{ab}- {1\over2}\,Rg_{ab}= \kappa T_{ab}\,,  \label{EFE}
\end{equation}
where $\kappa=8\pi G$ and $T_{ab}$ is the (symmetric) energy-momentum tensor of the matter. This is determined by the distribution of the local matter and satisfies the conservation law $\nabla^bT_{ab}=0$.

While the Ricci tensor monitors the local gravitational field, the Weyl tensor describes its long-range counterpart. In other words, the conformal curvature tensor represents the part of the spacetime curvature that is not determined by the local matter, but depends on the matter distribution at other points. This relates the Weyl field to tidal forces and gravitational waves. Further insight is obtained when $C_{abcd}$ is decomposed into its irreducible constituents. Given a 4-velocity congruence $u_a$, we have
\begin{equation}
C_{ab}{}^{cd}= 4\left(u_{[a}u^{[c}+h_{[a}{}^{[c}\right)E_{b]}{}^{d]}+ 2\varepsilon_{abe}u^{[c}H^{d]e}+ 2u_{[a}H_{b]e}\varepsilon^{cde}\,, \label{Weyl}
\end{equation}
with $E_{ab}=C_{acbd}u^cu^d$ and $H_{ab}=\varepsilon_a{}^{cd}C_{cdbe}u^e/2$ representing the electric and magnetic parts of the Weyl tensor (relative to the $u_a$-frame). Note that the magnetic component has no Newtonian analogue and for this reason is more closely (though not explicitly) associated with gravitational radiation. Also, both parts of the Weyl field are necessary if gravitational waves are to exist.

\subsection{Kinematics}\label{ssKs}
An advantage of 1+3 decomposition is its mathematical compactness and physical transparency, since every single one of the irreducible components involved has a very clear physical or geometrical interpretation. For example, the covariant derivative of the 4-velocity field splits as
\begin{equation}
\nabla_bu_a= {\rm D}_bu_a-A_au_b= {1\over3}\,\Theta h_{ab}+ \sigma_{ab}+ \omega_{ab}- A_au_b\,,  \label{Nbua}
\end{equation}
where ${\rm D}_a=h_a{}^b\nabla_b$ is the 3-D covariant derivative operator. The right-hand side of the above provides the irreducible kinematic variables that describe the observers' motion. These are the volume scalar, $\Theta=\nabla^au_a={\rm D}^au_a$, the shear and the vorticity tensors, $\sigma_{ab}={\rm D}_{\langle b}u_{a\rangle}$ and $\omega_{ab}={\rm D}_{[b}u_{a]}$ respectively, and the 4-acceleration vector, $A_a=\dot{u}_a$.\footnote{Round brackets denote symmetrisation and square ones antisymmetrisation. Angular brackets are used to describe the symmetric and trace-free part of orthogonally projected second rank tensors and vectors (e.g.~$\sigma_{ab}={\rm D}_{\langle b}u_{a\rangle}={\rm D}_{(b}u_{a)}-({\rm D}^cu_c/3)h_{ab}$ and $\dot{E}_{\langle a\rangle}=h_a{}^b\dot{E}_a$). Also, overdots indicate covariant differentiation with respect to proper time, namely along the fundamental worldlines (e.g.~$A_a=\dot{u}_a=u^b\nabla_bu_a$).} The latter represents non-gravitational forces and therefore vanishes when matter moves along geodesic worldlines. Also, a positive/negative volume scalar implies expansion/contraction for the fluid element, the shear modifies its shape and the vorticity changes its orientation. Note that, starting from the vorticity tensor, we may also define the vorticity vector by means of $\omega_a=\varepsilon_{abc}\omega^{bc}/2$, with $\varepsilon_{abc}$ representing the (totally antisymmetric) 3-D Levi-Civita tensor.

The motion of the fundamental observers is described by a set of three propagation equations supplemented by an equal number of constrains. These follow after applying the Ricci identities,
\begin{equation}
2\nabla_{[a}\nabla_{b]}u_c= R_{abcd}u^d\,,   \label{Ricci1}
\end{equation}
to the associated 4-velocity field. The timelike component of the above leads successively to Raychaudhuri's formula,
\begin{equation}
\dot{\Theta}= -{1\over3}\,\Theta^2- {1\over2}\,\kappa(\rho+3p)- 2\left(\sigma^2-\omega^2\right)+ {\rm D}^aA_a+ A^aA_a\,, \label{Ray1}
\end{equation}
to the shear-propagation equation
\begin{equation}
\dot{\sigma}_{\langle ab\rangle}= -{2\over3}\,\Theta \sigma_{ab}- \sigma_{c\langle a}\sigma^c{}_{b\rangle}- \omega_{\langle a}{}\omega_{b\rangle}+ {\rm D}_{\langle a}A_{b\rangle}+
A_{\langle a}A_{b\rangle}- E_{ab}+ {1\over2}\,\kappa\pi_{ab} \label{sheardot1}
\end{equation}
and to the evolution equation of the vorticity
\begin{equation}
\dot{\omega}_{\langle a\rangle}= -{2\over3}\,\Theta\omega_a- {1\over2}\,{\rm curl}A_{a}+ \sigma_{ab}\omega^b\,, \label{vortdot1}
\end{equation}
where $\sigma^2=\sigma_{ab}\sigma^{ab}/2$ and $\omega^2=\omega_{ab}\omega^{ab}/2=\omega_a\omega^a$. Expression (\ref{Ray1}) dictates the average kinematics of the fluid element, while (\ref{sheardot1}) and (\ref{vortdot1}) monitor the evolution of kinematically induced anisotropies and the rotational behaviour of the fluid respectively. For our purposes, the key equation is Raychaudhuri's formula, which governs the rate of the volume contraction/expansion of a self-gravitating medium. Negative terms in the right-nahd side of (\ref{Ray1}) assist the collapse (or inhibit the expansion), while positive ones resist contraction (or lead to expansion).

The spacelike component of Eq.~(\ref{Ricci1}) also splits in an analogous way. The result is expressed as a set of three (kinematic) constraints, given by
\begin{equation}
{\rm D}^b\sigma_{ab}= {2\over3}\,{\rm D}_a\Theta+ {\rm curl}\omega_a+ 2\varepsilon_{abc}A^b\omega^c- \kappa q_a\,, \label{shearcon}
\end{equation}
\begin{equation}
{\rm D}^a\omega_{a}= A^a\omega_a  \label{vortcon}
\end{equation}
and
\begin{equation}
H_{ab}= {\rm curl}\sigma_{ab}+ {\rm D}_{\langle a}\omega_{b\rangle}+ 2A_{\langle a}\omega_{b\rangle}\,, \label{Hcon}
\end{equation}
respectively. Together, relations (\ref{Ray1})-(\ref{Hcon}) provide a fully relativistic kinematic description of a general imperfect  fluid.

\subsection{Matter fields}\label{ssMFs}
The energy-momentum tensor of the matter also decomposes into its irreducible parts. When dealing with an imperfect fluid, for example, we have
\begin{equation}
T_{ab}=\rho u_au_b+ ph_{ab}+ 2q_{(a}u_{b)}+ \pi_{ab}\,.  \label{iTab}
\end{equation}
Here, $\rho=T_{ab}u^au^b$ is the energy density, $p=T_{ab}h^{ab}/3$ is the isotropic pressure, $q_a=-h_a{}^{b}T_{bc}u^c$ represents the energy flux and $\pi_{ab}=h_{\langle a}{}^{c}h_{b\rangle}{}^{d}T_{cd}$ the anisotropic pressure, all measured in the fundamental frame. In the case of a perfect fluid, $q_a$ and $\pi_{ab}$ vanish identically and the above reduces to
\begin{equation}
T_{ab}=\rho u_au_b+ ph_{ab}\,.  \label{pTab}
\end{equation}
with $p$ and $\rho$ related through the equation of state. For barotropic media, the latter takes the simple $p=p(\rho)$ form.

The nature of the matter fields also determines the conservation laws. Assuming a general imperfect fluid, the timelike component of the (twice contracted) Bianchi identities provides the energy conservation law
\begin{equation}
\dot{\rho}= -\Theta(\rho+p)- {\rm D}^aq_a- 2A^aq_a- \sigma^{ab}\pi_{ab}\,.  \label{ecl}
\end{equation}
Its spacelike counterpart, on the other hand, leads to the conservation of the momentum
\begin{equation}
(\rho+p)A_a= -{\rm D}_ap- \dot{q}_{\langle a\rangle}- {4\over3}\,\Theta q_a- \left(\sigma_{ab}+\omega_{ab}\right)q^b- {\rm D}^b\pi_{ab}- \pi_{ab}A^b\,.  \label{mcl1}
\end{equation}
As mentioned above, when dealing with perfect fluids, we may set $q_a=0=\pi_{ab}$. Then, expressions (\ref{ecl}) and (\ref{mcl1}) simplify considerably.

\section{Electromagnetic fields}\label{sEMFs}
The electromagnetic field is a source of energy density and, as such, it contributes to the right-hand side of the Einstein equations (see~(\ref{EFE})). In addition, the Maxwell field has its own evolution and constraint formulae, while it also obeys a separate set of conservation laws.

\subsection{Fluid description}
The electromagnetic field is invariantly described by the antisymmetric Faraday tensor ($F_{ab}$), which relative to our fundamental observers, splits as
\begin{equation}
F_{ab}=2u_{[a}E_{b]}+ \epsilon_{abc}B^c\label{Fab}
\end{equation}
where $E_{a}=F_{ab}u^b$ and $B_{a}=\epsilon_{abc}F^{bc}/2$ are the associated electric and magnetic fields respectively. The above decomposition also facilitates a fluid description of the Maxwell field, analogous to that seen in (\ref{iTab}). To be precise, given a 4-velocity field, the electromagnetic energy-momentum tensor splits as
\begin{equation}
T_{ab}^{(em)}= {1\over2}\left(E^2+B^2\right)u_au_b+ {1\over6}\left(E^2+B^2\right)h_{ab}+ 2\mathfrak{Q}_{(a}u_{b)}+ \Pi_{ab}\,,  \label{emTab}
\end{equation}
where $E^2=E_aE^a$ and $B^2=B_aB^a$ are the square magnitudes of the individual fields.\footnote{We use Heaviside-Lorentz units for the electromagnetic field throughout this article.} Also, $\mathfrak{Q}_a=\varepsilon_{abc}E^bB^c$ is the electromagnetic Poynting vector and $\Pi_{ab}=-E_{\langle a}E_{b\rangle}-B_{\langle a}B_{b\rangle}$. Comparing the above to Eq.~(\ref{iTab}), confirms that the Maxwell field corresponds to an imperfect fluid with energy density $\rho^{(em)}=(E^2+B^2)/2$, isotropic pressure $p^{(em)}=(E^2+B^2)/6$, energy flux $q^{em}_a=\mathfrak{Q}_a$ and anisotropic stresses given by $\Pi_{ab}$. Finally, decomposition (\ref{emTab}) ensures that $T_{a}^{(em)a}$=0, in agreement with the trace-free nature of the electromagnetic radiation.

\subsection{Maxwell's equation}\label{ssMEs}
Invariantly, Maxwell's equations are given by the set $\nabla_{[a}F_{bc]}=0$ and $\nabla^bF_{ab}=J_a$, with $J_a$ representing the electric 4-current. On using decomposition (\ref{Fab}), the above expressions decompose into a set of two propagation equations
\begin{eqnarray}
\dot{E}_{\langle a\rangle}&=& -{2\over3}\,\Theta E_a+ (\sigma_{ab}+\varepsilon_{abc}\omega^c)E^b+ \varepsilon_{abc}A^bB^c +{\rm curl}B_a-\mathfrak{J}_a\,,  \label{Ampere}\\
\dot{B}_{\langle a\rangle}&=& -{2\over3}\Theta B_a+ (\sigma_{ab}+\varepsilon_{abc}\omega^c)B^b- \varepsilon_{abc}A^b E^c- {\rm curl}E_a\,, \label{Faraday}
\end{eqnarray}
supplemented by the constraints
\begin{equation}
{\rm D}^aE_a+ 2\omega^aB_a= \mu  \label{Coulomb}
\end{equation}
and
\begin{equation}
{\rm D}^aB_a- 2\omega^aE_a= 0\,.  \label{Gauss}
\end{equation}
Note that ${\rm curl}B_a=\varepsilon_{abc}{\rm D}^bB^c$ by definition, with an analogous expression for ${\rm curl}E_a$. Also, $\mathfrak{J}_a$ represents the orthogonally projected (spatial) electric current and $\mu$ is the electric charge. The two are related to the 4-current by means of $\mathfrak{J}_a=h_a{}^bJ_b$ and $\mu=-J_au^a$ respectively. As a result, we may write
\begin{equation}
J_a= \mu u_a+ \mathfrak{J}_a\,.  \label{Ja}
\end{equation}

Relations (\ref{Ampere})-(\ref{Gauss}) are the covariant analogues of Ampere's law, Faraday's law, Coulomb's law and Gauss' law  respectively. A characteristic feature of these formulae is that they incorporate relative-motion effects, carried by the kinematic terms in their right-hand side. The first two terms in the right-hand side of the Ampere law, in particular, represent the magnetic field induced by the relative motion of its electric counterpart. Analogous terms are also seen in Faraday's law. Similarly, kinematical effects introduce an effective electric charge and an effective magnetic charge in Eqs.~(\ref{Coulomb}) and (\ref{Gauss}) respectively. As a result, the magnetic vector is generally not a solenoidal. Finally, the acceleration  terms in expressions (\ref{Ampere}) and (\ref{Faraday}) reflect the fact that spacetime is treated as a single, unified entity.

\subsection{Conservation laws}\label{ssCLs}
The electromagnetic stress-energy tensor is separately conserved (i.e.~$\nabla^bT_{ab}^{(em)}=0$). In the pesence of charged matter, however, one needs to substitute the total energy-momentum tensor into the Bianchi identities and then decompose the resulting relation. Then, using the full Einstein-Maxwell system, the timelike part of the aforementioned splitting reads
\begin{equation}
\dot{\rho}= -\Theta(\rho+p)- {\rm D}^aq_a- 2A^aq_a- \sigma^{ab}\pi_{ab}+ E^a\mathfrak{J}_a\,,  \label{emecl}
\end{equation}
and provides the energy conservation law. The spacelike component, on the other hand, leads to the conservation of the momentum
\begin{equation}
(\rho+p)A_a= -{\rm D}_ap- \dot{q}_{\langle a\rangle}- {4\over3}\,\Theta q_a- \left(\sigma_{ab}+\omega_{ab}\right)q^b- {\rm D}^b\pi_{ab}- \pi_{ab}A^b+ \mu E_a+ \varepsilon_{abc}\mathfrak{J}^bB^c\,.  \label{mcl2}
\end{equation}
Note that $q_a=0=\pi_{ab}$ for perfect fluids, in which case the right-hand sides of (\ref{emecl}) and (\ref{mcl2}) simplify considerably. Also, as we shall see below, the overall effect of the electromagnetic terms in the above depends on the electric properties of the matter.

Finally, when dealing with charged media, there is a an additional conservation law. This is related to the 4-current and is given by the constraint $\nabla^aJ_a=0$. The latter leads to the conservation law of the electric charge
\begin{equation}
\dot{\mu}= -\Theta\mu- {\rm D}^a\mathfrak{J}_a- A^a\mathfrak{J}_a\,, \label{ccl}
\end{equation}
expressed relative to a family of fundamental observers.

\subsection{Ohm's law}
The evolution and the effects of the Maxwell field depend on the electrical properties of the charged medium, which are described by Ohm's law. At the resistive MHD limit and in the frame of the fluid, the latter takes the covariant form~\cite{Gr,J}
\begin{equation}
J_a=\mu u_a+\varsigma E_a\,,  \label{Ohm1}
\end{equation}
with $\varsigma$ representing the (scalar) conductivity of the medium. Substituting decomposition (\ref{Ja}) into the left-hand side of (\ref{Ohm1}) and then projecting orthogonal to $u_a$, gives
\begin{equation}
\mathfrak{J}_a=\varsigma E_a\,.  \label{Ohm2}
\end{equation}
This, alternative form, of Ohm's law relates directly the 3-current to the $E$-field through the electrical properties of the host medium.

The two limiting cases are those of very poor and very high electrical conductivity, corresponding to $\varsigma\rightarrow0$ and $\varsigma\rightarrow\infty$ respectively. Following Eq.~(\ref{Ohm2}), the former has zero electric currents and the latter is characterised by a vanishing electric field. For any other intermediate situation, one needs a model for the conductivity of the matter.

\section{Charged collapse}\label{sCC}
Over the years, the gravitational contraction of charged matter (usually dust) in the presence of an electromagnetic (usually a purely electric) field has been considered by several authors from a variety of perspectives (see~\cite{Th1}-\cite{SA} for a representative though incomplete list). A common feature in a number of these studies, was the conclusion that charged dust could, under `favourable' circumstances, escape the ultimate collapse to a central singularity.

\subsection{Poorly conductive dust}\label{ssPCD}
Some of the aforementioned approaches are static, others are dynamical and, with very few exceptions, all are metric-based. Here, we will take a dynamical look into the question of charged collapse. We will also approach the problem from the 1+3-covariant point of view, instead of adopting metric-based techniques. This puts the Raychaudhuri equation (see relation (\ref{Ray1}) in \S~\ref{ssKs}) at the centre of our analysis.

Raychaudhuri's formula has played a fundamental role in gravitational-collapse studies and has been at the centre of all the singularity theorems. Expression (\ref{Ray1}) is the general one and can be adapted to any physical environment. When dealing with charged matter in the presence of an electromagnetic field, one needs to take into account the fact that the Maxwell field adds to the gravitational mass of the system and also incorporate the contribution of the Lorentz force to the momentum-conservation law. Assuming matter in the form of a pressureless perfect fluid, we may set $p$, $q_a$ and $\pi_{ab}$ to zero. Then, at the low conductivity limit, $\mathfrak{J}_a$ vanishes as well and the momentum-conservation law (see expression (\ref{mcl2}) in \S~\ref{ssCLs}) reduces to
\begin{equation}
A_a= \xi E_a\,,  \label{lconmc}
\end{equation}
where $\xi=\mu/\rho$ measures the specific charge density of the fluid.\footnote{To the best of our knowledge, all the available studies of charged collapse also ignore the presence of the electric currents. Seen it from our perspective, this corresponds to choosing poorly conductive environments.} Note that this ratio is time invariant in our case. Indeed, for a pressureless perfect fluid and for poor electrical conductivity, both $\rho$ and $\mu$ obey identical evolution laws (see Eqs.~(\ref{emecl}) and (\ref{ccl}) in \S~\ref{ssCLs}). We also underline the absence of any direct magnetic effects in the above relation, which results from the vanishing of the 3-currents.

In order to analyse the implications of a nonzero net charge for the average kinematics (i.e.~the volume expansion/contraction) of a self-gravitating fluid, we need to substitute (\ref{lconmc}) into the right-hand side of Eq.~(\ref{Ray1}). Then, on using Coulomb's law (see expression (\ref{Coulomb}) in \S~\ref{ssMEs}), the latter reads
\begin{eqnarray}
\dot{\Theta}&=& -{1\over3}\,\Theta^2- {1\over2}\,\kappa\left(\rho+E^2+B^2\right)- 2\left(\sigma^2-\omega^2\right)+ \xi^2\left(\rho+E^2\right) \nonumber\\ &&+E^a{\rm D}_a\xi- 2\xi\omega^aB_a\,.  \label{lconRay1}
\end{eqnarray}
In contrast to Eq.~(\ref{lconmc}), the above contains direct magnetic effects. These come from the $B$-field's contribution to the total energy density of the system and also to the effective charge (see relation (\ref{Coulomb}) in \S~\ref{ssMEs}). For our purposes, however, the key quantity is the (positive definite) fourth term in the right-hand side of Eq.~(\ref{lconRay1}). This term reflects the presence of a net electric charge and monitors the effect of the repulsive Coulomb forces on our self-gravitating system.

\subsection{Coulomb repulsion vs gravitational
attraction}\label{ssCRvsGA}
Assuming that spherically symmetry and spatial homogeneity are largely preserved, we may ignore the shear and the vorticity, as well as variations in the spatial distribution of the charge density. In that case Raychaudhuri's equation simplifies to
\begin{equation}
\dot{\Theta}+ {1\over3}\,\Theta^2= -{1\over2}\,\kappa\left(\rho+E^2+B^2\right)+ \xi^2\left(\rho+E^2\right)\,.  \label{lconRay2}
\end{equation}
The above describes the average volume contraction/expansion of an almost spherically symmetric, charged perfect fluid of low electrical conductivity.\footnote{In the case of exact spherical symmetry the magnetic field vanishes in the frame of the fluid, while only the radial component of its electric counterpart survives (e.g.~see~\cite{KB,LL}).} Put another way, Eq.~(\ref{lconRay2}) monitors the average separation between the (timelike) worldlines of two neighbouring observers that move along with the matter. We remind the reader that, in the presence of a nonzero net charge, these worldlines are no longer geodesics.

In the right-hand side of (\ref{lconRay2}) one can clearly see the competing action of the gravitational and the Coulomb forces and how these affect the kinematics of the charged fluid. The first term, in particular, shows how the Maxwell field increases the overall pull of gravity by adding to the total energy density of the system. Alternatively, one might say that the electric charge contributes to the total gravitational mass, an effect occasionally referred to as `charge regeneration'~\cite{B}. The second term, on the other hand, demonstrates how the repulsive Coulomb forces act against gravity. Note that this stress is partly Newtonian and partly relativistic. The Newtonian component is $\xi^2\rho$ and derives from the 3-divergence of the 4-acceleration in Eq.~(\ref{Ray1}), which also appears in non-relativistic treatments~\cite{El2,ST}. The relativistic effects, on the other hand, are encoded in the $\xi^2E^2$ stress. The latter comes from the square of the 4-acceleration in the right-hand side of the Raychaudhuri equation and has no Newtonian counterpart~\cite{El2,ST}. All these mean that the electric field contributes equally to the total gravitational field and to the Coulomb forces. In contrast, the magnetic field only adds to the overall pull of gravity. This difference reflects the absence of magnetic monopoles.\footnote{If magnetic monopoles were also included into the scheme, their presence would have led to a Coulomb-like repulsive stress in the right-hand side of Eq.~(\ref{lconRay2}), analogous to the one produced by the electric charges.}

It is a very well known result that, as long as the right-hand side of Eq.~(\ref{lconRay2}) remains non-positive, an initially converging pair of worldlines will focus within finite proper time. This implies that, within our approximation scheme, the Coulomb forces can stop these worldlines from focusing (and therefore prevent caustic formation) if
\begin{equation}
\xi^2\left(\rho+E^2\right)> {1\over2}\,\kappa\left(\rho+E^2+B^2\right)\,.  \label{lconcon1}
\end{equation}
When the above condition holds, the worldlines of a contracting, nearly spherically symmetric, poorly conductive, pressureless, charged fluid will not focus. Moreover, in the absence of the magnetic field, as it happens in the case of exact spherical symmetry for example, the above reduces to
\begin{equation}
\xi> \sqrt{{1\over2}\,\kappa}= \sqrt{{1\over2}}\,M_{Pl}^{-1}\,,  \label{lconcon2}
\end{equation}
having set $\kappa=M_{Pl}^{-2}$ (with $M_{Pl}$ representing the Planck mass). Note that the last condition has also been discussed, as a requirement for avoiding both the initial and shell-crossing singularities during the spherically symmetric collapse of charged dust, in the metric based approach of~\cite{KB}. Analogous conclusions, relating the charge-to-matter ratio with the stability (static or dynamical) of a spherically symmetric charged medium against gravitational contraction, were also reached in the earlier treatments of~\cite{B,R1}.

The final form of (\ref{lconcon1}) also depends on the strength of the electromagnetic components relative to the matter content. Assuming that $E^2$, $B^2\ll\rho$, we arrive again at $\xi>\sqrt{\kappa/2}= M_{Pl}^{-1}/\sqrt{2}$. When  $E^2\simeq B^2\gg\rho$, on the other hand, condition (\ref{lconcon1}) recasts into $\xi>\sqrt{\kappa}=M_{Pl}^{-1}$. Finally, for $E^2\simeq B^2\simeq\rho$, expression (\ref{lconcon1}) takes the form $\xi>\sqrt{3\kappa/4}=\sqrt{3/4}M_{Pl}^{-1}$. In all these cases the repulsive Coulomb forces will be strong enough to counterbalance the gravitational field and prevent the (non-geodesic) worldlines of the charged particles from focusing.

Before closing this section, we should also briefly discuss the subtle role played by possible inhomogeneities in the spatial distribution of the electric charge. These are incorporated in the second-last term of Eq.~(\ref{lconRay1}) and their effects have so far been bypassed. When included into expression (\ref{lconRay2}), the latter reads
\begin{equation}
\dot{\Theta}+ {1\over3}\,\Theta^2= -{1\over2}\,\kappa\left(\rho+E^2+B^2\right)+ \xi^2\left(\rho+E^2\right)+ E^a{\rm D}_a\xi\,,  \label{lconRay3}
\end{equation}
where $E^a{\rm D}_a\xi\gtrless0$ in general. Depending on its sign, the inhomogeneous term in the right-hand of the above will either strengthen or weaken conditions (\ref{lconcon1}) and (\ref{lconcon2}). In particular, when
\begin{equation}
E^a{\rm D}_a\xi= \xi\left({1\over\mu}\,E^a{\rm D}_a\mu -{1\over\rho}\,E^a{\rm D}_a\rho\right)> 0\,,  \label{lconcin3}
\end{equation}
the spatial gradients of the specific charge will assist the Coulomb repulsion and therefore hamper the collapse further. In the opposite case, the situation is reversed and the inhomogeneous term in the right-hand side of (\ref{lconRay3}) enhances the overall gravitational pull. This versatility in the effect of the charge gradients has been noted in earlier studies as well~\cite{Be,PHDMcLS}. The reader is particularly referred to~\cite{PHDMcLS} for an alternative, metric-based, discussion of the matter.

\subsection{Highly conductive dust}\label{ssHCD}
The assumption of a charged-matter distribution in the absence of magnetic fields and electric currents is probably a highly idealised one. Put another way, poorly conductive environments offer a rather unlikely description of stellar interiors. Strong magnetic fields and high electrical conductivity sound much more likely.

When dealing with highly conductive matter, Ohm's law (see Eq.~(\ref{Ohm2})) guarantees that the electric fields vanish in the frame of the fluid. At the same time, the finite currents keep the magnetic field lines frozen in with the matter~\cite{P,M}. This is the familiar ideal-MHD approximation, which has long been used in studies of magnetised plasmas in both astrophysics and cosmology. Here, we will also assume overall charge neutrality. In such environments, Maxwell's formulae (see relations (\ref{Ampere})-(\ref{Gauss})) reduce to one propagation equation and three constrains. More specifically, Ampere's law leads to the magnetic induction equation
\begin{equation}
\dot{B}_{\langle a\rangle}= -\frac{2}{3}\,\Theta B_a+ (\sigma_{ab}+\epsilon_{abc}\omega^c)B^b\,,  \label{MHDM1}
\end{equation}
while Faraday's, Coulomb's and Gauss' laws reduce to the constrains
\begin{equation}
\varepsilon_{abc}A^bB^c+ \mathrm{curl}B_a= \mathfrak{J}_a\,, \hspace{20mm} \omega^aB_a= 0 \hspace{10mm} {\rm and} \hspace{10mm} {\rm D}^aB_a= 0\,,  \label{MHDM2}
\end{equation}
respectively. Note that Eq.~(\ref{MHDM1}) guarantees that the magnetic lines always connect the same particles (i.e.~that the $B$-field is frozen-in with the matter).

The kinematics of a highly conductive pressureless fluid are still governed by the set (\ref{Ray1})-(\ref{Hcon}), once the latter is adapted to the specifics of the MHD limit. In particular, Raychaudhuri's equation reads
\begin{equation}
\dot{\Theta}= -{1\over3}\,\Theta^2- {1\over2}\,\kappa\left(\rho+B^2\right)- 2\left(\sigma^2-\omega^2\right)+ {\rm D}^aA_a+ A^aA_a\,,  \label{MHDRay1}
\end{equation}
with the 4-acceleration given by the momentum-conservation law, which has now changed to
\begin{equation}
\left(\rho+B^2\right)A_a= -\epsilon_{abc}B^b{\rm curl}B^c= -{1\over2}\,{\rm D}_aB^2+ B^b{\rm D}_bB_a\,,  \label{MHDmc2}
\end{equation}
The above set monitors the average contraction/expansion of a self-gravitating, highly conductive, pressureless fluid in a magnetised environment. At this point, we would like to draw the readers attention to the second equality in Eq.~(\ref{MHDmc2}), where the Lorentz force has been split into its pressure and tension components. The former comes from the positive pressure that the $B$-field exerts orthogonal to its own direction and demonstrates the tendency of the magnetic forcelines to push each other apart. The latter reflects the negative pressure, namely the tension, that the field exerts along its own direction. This stress manifests a unique feature of magnetic fields: the elasticity of their forcelines and their tendency to remain `straight'~\cite{P,M}.

\subsection{Magnetic tension effects}\label{ssMTEs}
Because of their elastic properties, the field lines react to any attempt that distorts them from equilibrium. In standard Newtonian MHD, the agents causing such deformations are ordinary charged particles. The field's reaction then triggers a variety of tension stresses, which in turn lead to a range of generally well studied effects. When General Relativity is involved, however, there is an additional agent that can affect the distribution and the topology of the magnetic forcelines. This new agent is the curvature of the spacetime itself.

The vector nature of magnetic fields means that they `feel' the curvature of the space via the Ricci identities, as well as through Einsten's equations. The former are a purely geometrical set of relations that have played a fundamental role in the mathematical formulation of general relativity, since they essentially provide the definition of spacetime curvature. Applied to the magnetic vector, the Ricci identities read
\begin{equation}
2\nabla_{[a}\nabla_{b]}B_c= R_{abcd}B^d\,,  \label{Ricci2}
\end{equation}
with $R_{abcd}$ representing the Riemann tensor of the spacetime (see Eq.~(\ref{Riemann}) in \S~\ref{ssGF}). Starting from the above, one can obtain the so-called 3-Ricci identity
\begin{equation}
2{\rm D}_{[a}{\rm D}_{b]}B_c= \mathcal{R}_{abcd}B^d- 2\omega_{ab}\dot{B}_{\langle c\rangle}\,,  \label{3Ricci}
\end{equation}
which now involves $\mathcal{R}_{abcd}$, namely the Riemann tensor of the 3-dimensional hypersurfaces orthogonal to our fundamental observers. Note the vorticity term in the right-hand side of the above, which reflects the fact that in a rotating spacetime the observers' worldlines are not hypersurface orthogonal. The Ricci identities demonstrate, mathematically, how distortions in the geometry of the host spacetime translate into magnetic-line deformations. These in turn lead to a range of tension stresses that are purely geometrical in origin. Next, we will concentrate on the implications of such tension stresses for the gravitational collapse of a magnetised fluid.

For our purposes, there is no real loss of generality in assuming (almost) zero rotation and (nearly) spatially homogeneous distributions for both the matter and the magnetic energy densities. The latter assumption, in particular, allows us ignore the pressure part of the Lorentz force (see Eq.~(\ref{MHDmc2})) and focus on its tension component. In that case, the momentum-conservation law reduces to $(\rho+B^2)A_a=B^b{\rm D}_bB_a$. Substituted this expression into the right-hand side of Eq.~(\ref{MHDRay1}) and using relations (\ref{MHDM2}c), (\ref{3Ricci}), gives
\begin{equation}
\dot{\Theta}+ {1\over3}\,\Theta^2= -R_{ab}u^au^b+ c_a^2\mathcal{R}_{ab}n^an^b- 2\left(\sigma^2-\sigma_B^2\right)\,,  \label{MHDRay2}
\end{equation}
where $R_{ab}$ and $\mathcal{R}_{ab}$ are the 4-dimensional and the 3-dimensional Ricci tensors respectively $c_a^2=B^2/(\rho+B^2)$ is the square of the Alfv\'en speed and $n_a$ is the unit vector along the direction of the magnetic field. Also note that $2\sigma_B^2= ({\rm D}_{\langle a}B_{b\rangle})^2/(\rho+B^2)$ and we have dropped the last term of Eq.~(\ref{MHDRay1}). That quantity, namely the square of the 4-acceleration -- $A^aA_a$, always resists gravitational contraction and its absence will not weaken our argument (see below).

The first term in the right-hand side of Eq.~(\ref{MHDRay2}) provides the total gravitational mass of the system, with $R_{ab}u^au^b=\kappa(\rho+B^2)/2>0$ in our case. Typically, this is the main quantity that drives the contraction, with the shear adding to the overall pull of gravity. The magnetic terms in (\ref{MHDRay2}) are due to the field's tension and are triggered by the deformation of its forcelines. Both terms represent restoring stresses, which act against the agent that caused the line distortion in the first place. The magneto-shear, for example, is triggered by shearing effects. This explains why $\sigma_B$ acts against shear proper in Eq.~(\ref{MHDRay2}). Note that this term is non-relativistic in origin, since it also appears in purely Newtonian treatments~\cite{ST}.\footnote{When rotation is also included into the system, it adds a tension stress to the right-hand of \ref{MHDRay2}) that may be seen as the magnetic analogue of vorticity proper~\cite{T1}. This stress is triggered by the twisting of the field lines, which trapped within the rotating fluid. In analogy with the magneto-shear term in Eq.~(\ref{MHDRay2}), the magnetic vorticity stress acts against its kinematic counterpart, an effect that is sometimes referred to as `magnetic braking'.}

\subsection{Magnetic tension vs gravitational
attraction}\label{ssMTvsGA}
Of all the quantities in the right-hand side of (\ref{MHDRay2}), the most unusual and intriguing is the second. This also represents a tension stress, but it is purely relativistic in nature. The difference with the other tension stresses is that, now, the agent responsible for the deformation of the magnetic field lines is the geometry of the 3-space. In particular, spatial curvature distortions affect the topology of the magnetic forcelines, giving rise to the mageto-curvature tension stress seen in Eq.~(\ref{MHDRay2}). As mentioned earlier, the whole interaction is mathematically monitored via the Ricci identities. Not surprisingly, the magnitude of the tension stress depends on the relative strength of the $B$-field and on the amount of the line deformation. The latter is determined by the sum $\mathcal{R}_{ab}n^an^b$, which represents the 3-curvature distortion along the direction of the field lines. What is surprising, is that the effect of the magneto-geometrical stress in (\ref{MHDRay2}) depends on the sign of the curvature deformation. Thus, the aforementioned stress will resist the collapse when the scalar $\mathcal{R}_{ab}n^an^b$ is positive, something that is generally expected in all cases of realistic stellar collapse. Moreover, the strength of the tension stress increases with growing curvature distortion, just like the resisting power of a spring grows when it is compressed back to its limit. All these mean that, at least in principle, it is conceivable that the aforementioned stress can grow strong enough to halt the contraction and therefore prevent the particle worldlines form focussing.\footnote{Recall that, in the presence of a non-zero Lorentz force, the worldlines in question are no longer geodesics.} Whether this happens or, depends on a number of parameters. Perhaps the decisive factor is the balance between the two geometrical terms in the right-hand side of Eq.~(\ref{MHDRay2}). Whether, in particular, the tension stress can outbalance the gravitational pull of the matter. In what follows we will take a closer look into this possibility.

Suppose that at some time during the collapse of a magnetised fluid the magneto-geometrical tension stress has grown strong enough, so that
\begin{equation}
c_a^2\mathcal{R}_{ab}n^an^b>R_{ab}u^au^b\,.  \label{MHDcon1}
\end{equation}
Then, substituting the Gauss-Codacci equation (e.g.~see expression (1.3.39) in~\cite{TCM} -- with zero rotation in our case) into the left-hand side of the above, we obtain
\begin{equation}
{2\over3}\,c_a^2\left(\kappa\rho-{1\over3}\,\Theta^2\right)+ c_a^2\left(E_{ab}-{1\over3}\,\Theta\sigma_{ab}+ \sigma_{ca}\sigma^c{}_b\right)n^an^b> {1\over2}\,\kappa\left(\rho+B^2\right)\,.  \label{MHDcon2}
\end{equation}
The first of the two parentheses on the left represents the isotropic part of the tension stress and the second the anisotropic. In particular, $E_{ab}$ represents the electric component of the Weyl tensor, which is known to describe the free gravitational field. We now introduce the auxiliary variables $\beta=B^2/\rho$ and $\Omega_{\rho}= 3\kappa\rho/\Theta^2$. The former measures the relative strength of the $B$-field and the latter corresponds to the familiar density parameter typically associated with the FRW universes. Then, condition (\ref{MHDcon2}) takes the form
\begin{equation}
\mathfrak{R}_{ab}n^an^b>{1\over3}\,\kappa\rho\left(1+{3\over2}\,\beta +{3\over2}\,\beta^{-1}+2\Omega_{\rho}^{-1}\right)>0\,,  \label{MHDcon3}
\end{equation}
where $\mathfrak{R}_{ab}=E_{ab}-(\Theta/3)\sigma_{ab}+ \sigma_{ca}\sigma^c{}_b$. The fact that the scalar $\mathfrak{R}_{ab}n^an^b$ is nonzero implies that the collapse must be anisotropic, if the tension stress is to outbalance the gravitational pull of the matter. Calculating the required amount of anisotropy is a rather involved process, as it depends on a number of (essentially free) parameters. Nevertheless, some general estimates are still possible.

The first, qualitative, result comes from the dependence of $\mathfrak{R}_{ab}n^an^b$ on the electric Weyl tensor and also on the shear. This dependence suggests that the tension stresses are more likely to outbalance the pull of gravity in situations where the Weyl part of the gravitational field dominates over its Ricci counterpart. More quantitative estimates can be obtained by assuming that $\Omega_{\rho}\gg1$ during the advanced stages of the contraction, which sounds plausible for most realistic collapse scenarios. Then, following (\ref{MHDcon3}), the required amount of anisotropy depends entirely on the relative strength of the $B$-field (i.e.~on the $\beta$-parameter). Assuming that the two sources are equally strong (i.e.~setting $\beta\sim1$), we find that condition (\ref{MHDcon3}) is satisfied if $\mathfrak{R}_{ab}n^an^b\gtrsim\rho$. This changes to $\mathfrak{R}_{ab}n^an^b\gtrsim B^2$ when the magnetic field dominates (i.e.~for $\beta\gg1$). In the opposite case, namely when $\beta\ll1$, expression (\ref{MHDcon3}) leads to $\mathfrak{R}_{ab}n^an^b\gg\rho$. Overall, when the magnetic field is equally strong or stronger than the matter, the tension stresses can outbalance the gravitational pull of the matter fields provided the collapse is sufficiently (though not necessarily highly) anisotropic.\footnote{Even when the magnetic field dominates (i.e.~when $B^2\gg\rho$), the total fluid satisfies the three energy conditions (i.e.~weak, strong and dominant). One can easily verify this by recalling that $\rho_B=B^2/2$, while $p_B=-B^2/2$ along the direction of the field lines (i.e.~where the magnetic tension acts) and $p_B=B^2/2$ orthogonal to them.} This changes when the $B$-field is weak. Then, the magneto-geometrical tension stress will remain subdominant unless the collapse is highly anisotropic. Note, however, that the anisotropy requirements may weaken (perhaps significantly), if the effect of the repulsive $A_aA^a$-stress in the right-hand side of the Raychaudhuri equation (see expression (\ref{MHDRay1})) are also accounted for.

\section{Discussion}
The gravitational contraction of charged matter and the implications of electromagnetism for the fate of the collapse have been studied by many authors, from a variety of perspectives and by means of different methods. A number of these studies raised the possibility of a `bounce', which sometimes was caused by the presence of a net charge, while in others it was triggered by the presence of the magnetic field. The available approaches are static as well as dynamical and in almost all cases they employ metric-based techniques.

In this work we have adopted covariant methods to pursue a dynamical study of charged collapse. This has put the Raychaudhuri equation at the centre of our analysis. We have considered two cases that correspond to two opposite environments. The first assumes poorly conductive matter and looks at the effects of net charges, and on those of the resulting repulsive Coulomb forces, upon gravitational contraction. The second considers magnetised environments of high electrical conductivity and focuses on the role of the magnetic tension. In the first case we found that, to a very large extent, the overall outcome of the collapse depends on the value of the specific charge (i.e.~on the ratio between the charge and the energy densities) of the medium. When the charge density dominates, the Coulomb forces become stronger that the gravitational pull of the matter and the contraction stops. In the opposite case, the collapse proceeds essentially unimpeded by the Coulomb stresses. Our results agree both qualitatively and quantitatively with those obtained by earlier, primarily metric-based, studies.

Highly conductive environments probably offer a better approximation of realistic gravitational contraction. This alternative corresponds to the familiar ideal-MHD limit, where there are no electric fields and the currents keep the magnetic lines frozen-in with the fluid. In the absence of net charges, we found that the main resistance against gravitational collapse is probably the one coming from the magnetic tension. The latter reflects the elasticity of the field lines and their tendency to react against anything that distorts them from equilibrium. Assuming that General Relativity is the correct theory of gravity, one of the agents leading to such line-deformations is the curvature of the space. This means that magnetic forcelines trapped within a collapsing medium are being increasingly deformed by the ever growing gravitational field. These deformations inevitably lead to magneto-geometrical tension stresses that resist further contraction. A close analogy is that of a spring's tension, which grows stronger as the latter is compressed back to its limit. It is therefore conceivable that, under certain circumstances, the aforementioned elastic stresses may become strong enough to outbalance the pull of gravity and bring the collapse to a halt. Our analysis shows that, if something like that is to happen, it would probably need some degree of anisotropy. Although the exact amount of the required anisotropy is rather difficult to estimate at the moment, Weyl-curvature dominated spacetimes may offer a suitable host to test this hypothesis.


\begin{thebibliography}{99}
\bibitem{B} W.B. Bonnor, Z. Phys. \textbf{160}, 59 (1960).
\bibitem{M1} M.A. Melvin, Phys. Lett. \textbf{8}, 65 (1964).
\bibitem{M2} M.A. Melvin, Phys. Rev. B \textbf{139}, 225 (1965).
\bibitem{Th1} K.S. Thorne, Phys. Rev. B {\bf 138}, 251 (1965).
\bibitem{Th2} K.S. Thorne, Phys. Rev. B {\bf 139}, 244 (1965).
\bibitem{N} I.D. Novikov, Soviet Astr. {\bf 10}, 731 (1967).
\bibitem{dlCI} V. de la Cruz and W. Israel, Nuovo Cimento A \textbf {51}, 744 (1967).
\bibitem{Be} J.D. Bekenstein, Phys. Rev. D \textbf{4}, 2185 (1971).
\bibitem{R1} A.K. Raychaudhuri, Ann. Inst. Henri Poincar\`e \textbf{22}, 229 (1975).
\bibitem{AP} H. Ardavan and M.H. Partovi, Phys. Rev. D \textbf{16}, 1664 (1977).
\bibitem{O} A. Ori, Phys. Rev. D {\bf 44}, 2278 (1991).
\bibitem{dFYF} F. de Felice, Y. Yu and J. Fang, Mon. Not. R. Astron. Soc. \textbf{277}, L17 (1995).
\bibitem{dFSY} F. de Felice, L. Siming and Y. Yunqiang, Class. Quantum Grav. \textbf{16}, 1669 (1999).
\bibitem{REMLZ} S. Ray, A.L. Espindola, M. Malheiro, J.P.S. Lemos and V.T. Zanchin, Phys. Rev. D. \textbf{68}, 084004 (2003).
\bibitem{G} C.R. Ghezzi, Phys. Rev. D \textbf{72} 104017 (2005).
\bibitem{KB} A. Krasinski and K. Bolejko, Phys. Rev. D \textbf{73}, 124033 (2006).
\bibitem{T1} C.G. Tsagas, Class. Quant. Grav. {\bf 23}, 4323 (2006).
\bibitem{LL} P.D. Lansky and A.W.C. Lun, Phys. Rev. D \textbf{75}, 104010 (2007).
\bibitem{PHDMcLS} A. Di Prisco, L. Herrera, G. Le Denmat, M.A.H. MacCallum and N.O. Santos, Phys. Rev. D \textbf{76}, 064017 (2007).
\bibitem{SA} M. Sharif and G. Abbas, Astrophys. Space Sci. \textbf{327}, 285 (2010).
\bibitem{LZ} J.P.S. Lemos, V.T. Zanchin, Phys. Rev.D \textbf{81}, 124016 (2010).
\bibitem{R2} A.K. Raychaudhuri, Phys. Rev. D \textbf{90}, 1123 (1955).
\bibitem{E} J. Ehlers, Pramana \textbf{69}, 7 (2007).
\bibitem{El1} G.F.R. Ellis, Pramana \textbf{69}, 15 (2007).
\bibitem{D} N. Dadhich, Pramana \textbf{69}, 23 (2007).
\bibitem{KS} S. Kar and S. Sengupta, Pramana \textbf{69}, 49 (2007).
\bibitem{TCM} C.G. Tsagas, A. Challinor and R. Maartens, Phys. Rep.  {\bf 465}, 61 (2008).
\bibitem{Gr} P.J. Greenberg, Astrophys. J. \textbf{164}, 589 (1971).
\bibitem{J} J.D. Jackson, Classical Electromagnetism (Wiley, New York, 1975).
\bibitem{El2} G.F.R. Ellis, Mon. Not. R. Astron. Soc. \textbf{243}, 509 (1990).
\bibitem{ST} N.K. Spyrou and C.G. Tsagas, Mon. Not. R. Astron. Soc. \textbf{388}, 187 (2008).
\bibitem{P} E.N. Parker, Cosmical Magnetic Fields (Oxford University Press, Oxford, 1979).
\bibitem{M} L. Mestel, Stellar Magnetism (Oxford University Press, Oxford, 1999).
\end{thebibliography}
\end{document}